\documentclass[%
 reprint,amsmath,amssymb,aps,floatfix,showkeys,showpacs,
]{revtex4-1}

\usepackage[T1]{fontenc}
\usepackage[utf8]{inputenc}
\usepackage{graphicx,colordvi,color}
\usepackage{dcolumn}
\usepackage{bm}

\begin{document}

\title{Extended KdV equation for the case of uneven bottom}

\author{Piotr Rozmej}
 \email{P.Rozmej@if.uz.zgora.pl}\thanks{Corresponding author}
\affiliation{Institute of Physics, Faculty of Physics and Astronomy \\
University of Zielona G\'ora, Szafrana 4a, 65-246 Zielona G\'ora, Poland}

\author{Anna Karczewska}
 \email{A.Karczewska@wmie.uz.zgora.pl}
\affiliation{Faculty of Mathematics, Computer Science and Econometrics\\ University of Zielona G\'ora, Szafrana 4a, 65-246 Zielona G\'ora, Poland}

\date{\today} 

\begin{abstract}
We derived consistently, according to the second order perturbation approach, the extended KdV equation for an uneven bottom for the case of $\alpha=O(\beta)$ and $\delta=O(\beta^2)$. This equation can be obtained only when the bottom is given by a piecewise linear function. 
For the case of $\alpha=O(\beta)$ and $\delta=O(\beta)$ a unidirectional wave equation is derived in first order approach with the same limitation for the bottom profile.
\end{abstract}



\maketitle

\section{Introduction} \label{intro}

In 2014, with our co-workers, we derived the nonlinear second order wave equation for shallow water problem with uneven bottom \cite{KRR,KRI}. In these papers, besides standard small parameters $\alpha=\frac{a}{h}$ and $\beta=\left(\frac{h}{l}\right)^2$ we introduced the third one defined as $\delta=\frac{a_h}{h}$. In these definitions $a$ denotes the wave amplitude, $h$ the average water depth, $l$ the average wavelength and $a_h$ the amplitude of the bottom variations. We considered the case of $\alpha=O(\beta)$ and $\delta=O(\beta)$, that is, when all three small parameters are of the same order.
Then, with standard assumptions for incompressible, inviscid fluid and irrotational motion, we applied the second-order perturbation approach to the set of Eulerian equations. This set, written in nondimensional variables has the following form (see, e.g., Eqs.\ (2)-(5) in \cite{KRI}) 
\begin{eqnarray} \label{2BS}
\beta \phi_{xx}+\phi_{zz} & = &0, \\ \label{4BS}
\eta_t+\alpha\phi_x\eta_x-\frac{1}{\beta}\phi_z & = &0,~\mbox{for}~  z = 1+\alpha \eta\\  \label{5BS}
\phi_t+\frac{1}{2}\alpha \phi_x^2+\frac{1}{2}\frac{\alpha}{\beta}\phi_z^2 +\eta 
& = &0,~\mbox{for}~  z = 1+\alpha \eta \\ \label{6BS}
\phi_z-\beta\delta\left( h_x\,\phi_x\right) & = & 0,~\mbox{for}~ z=\delta h(x) .
\end{eqnarray}
Equation (\ref{2BS}) is the Laplace equation valid for the whole volume of the fluid. Equations (\ref{4BS}) and (\ref{5BS}) are so-called kinematic and dynamic boundary conditions at the surface, respectively. The equation  (\ref{6BS}) represents the boundary condition at the non-flat bottom. For abbreviation all subscripts denote the partial derivatives with respect to particular variables, i.e.\ $\phi_{x}\equiv \frac{\partial \phi}{\partial x}, \eta_{2x}\equiv \frac{\partial^2 \eta}{\partial x^2}$ and so on. 

For the flat bottom, the boundary condition at the bottom is $\phi_z=0$. In this case, the perturbation approach of the first order with respect to small parameters leads to the famous Korteweg-de Vries equation \cite{kdv}
\begin{equation}\label{kdv}
\eta_t +\eta_x+\alpha\frac{3}{2}\eta\eta_x +\beta \frac{1}{6}\eta_{3x}=0 .
\end{equation}
In second order, Marchant and Smyth obtained in 1990 the \emph{extended KdV} equation (called also KdV2) of the form \cite{MS90}
\begin{align} \label{etaabd}  
\eta_t +\eta_x&  + \alpha \frac{3}{2}\eta\eta_x +\beta\frac{1}{6} \eta_{3x}  + \alpha^2\left(\!-\frac{3}{8}\eta^2\eta_x \!\right)  \\ &  + \alpha\beta\left(\!\frac{23}{24}\eta_x\eta_{2x}\!+\!\frac{5}{12}\eta\eta_{3x}\! \right)+\beta^2\frac{19}{360}\eta_{5x}   = 0. \nonumber \end{align}

In \cite{KRR,KRI} we tried to extend the second-order approach to the case $\delta\ne 0$ of the non-flat bottom. Then the equation (\ref{6BS}), limited to the third order, allows us to express the velocity potential in the form \cite[Eq.~(7)]{KRI}
\begin{align} \label{pot1} 
\phi   & =    \phi^{(0)}_{x}\!+\!z\beta\delta \left(\! h  \phi^{(0)}_x\! \right)_{x}\!\! -\!\frac{1}{2}z^2 \beta \, \phi^{(0)}_{2x}  \!-\!\frac{1}{6}z^3 \beta^2\delta \left(\! h\phi^{(0)}_x \! \right)_{3x}  \\ & +\! \frac{1}{24}z^4 \beta^2 \phi^{(0)}_{4x}  \!+\! \frac{1}{120}z^5 \beta^3\delta\left(\! h \phi^{(0)}_x \! \right)_{5x} \! \!-\!  \frac{1}{720}z^6 \beta^3 \phi^{(0)}_{6x}+\ldots \nonumber  
\end{align}
Inserting (\ref{pot1}) into (\ref{4BS}) and (\ref{5BS}) and retaining only terms up to second-order one obtains the second-order Boussinesq's system
\begin{align} \label{4hx}
  \eta_t + w_x & + \alpha(\eta w)_x-\frac{1}{6}\beta w_{3x}-\frac{1}{2} \alpha\beta (\eta w_{2x})_x  \\ & -\frac{1}{120}\beta^2 w_{5x} -
 \delta (hw)_x +\frac{1}{2}\beta\delta (hw)_{3x} =0, \nonumber \\ \label{5hx}
w_t + \eta_x & + \alpha w w_x -\frac{1}{2}\beta\, w_{2xt} + \frac{1}{24}\beta^2\, w_{4xt} + \beta\delta\, (h w_t)_{2x}  \nonumber  \\ &
+ \frac{1}{2} \alpha\beta\left[-2(\eta w_{xt})_x +  w_x w_{2x} - w w_{3x} \right]  =0 . 
\end{align} 
This set of Boussinesq's equations is correct. 

Recently, it was pointed out in \cite{Burde} that our next steps,
performed in \cite{KRR,KRI} and leading to the KdV2 equation for uneven bottom were inconsistent, and therefore the derived equation \cite[Eq.~(18)]{KRI} bears no relevant solution to the problem considered.

We agree with this criticism. We derived our equation \cite[Eq.~(18)]{KRI} in good faith. However, using different notations for small parameters $\alpha,\beta,\delta$ we did not recognize the proper order of terms related to the bottom function.

The next parts of this article contain the following results. 
\begin{itemize}
\item The Boussinesq's system (\ref{4hx})-(\ref{5hx}) cannot be reduced (for arbitrary shape of the bottom function) to a single KdV-type equation even in the first order. In consequence, the same is true for any higher order equations for the case of $\alpha=O(\beta)$, $\delta=O(\beta)$.
We show this in Section \ref{Sabd}.
The author of \cite{Burde} found that first order KdV-type equation for the case $\alpha=O(\beta)$, $\delta=O(\beta)$ can be derived for a very special case of a linear bottom function $h=kx$.
We showed that this result is correct for arbitrary piecewise linear bottom function.

\item 
For the case of $\alpha=O(\beta)$, $\delta=O(\beta^2)$ the appropriate Boussinesq's system (\ref{4HX})-(\ref{5HX}) cannot be reduced (for arbitrary shape of the bottom function) to a single KdV2-type equation. 
Similarly, as in the previous case, the KdV2-type equation can be derived only for a piecewise linear bottom function.

\item In Section \ref{secN} we test motion of solitons over the uneven bottom of the trapezoidal shape. This bottom function is piecewise linear. Two cases, a bump, and a well are tested. Initial conditions are taken as the KdV solitons for the case of $\alpha=O(\beta)$ and $\delta=O(\beta)$ and as the KdV2 solitons
for the case of $\alpha=O(\beta)$ and $\delta=O(\beta^2)$.
\end{itemize}

\section{(Non)existence of wave equation for the case of  \boldmath$\alpha=O(\beta)$ and  \boldmath$\delta=O(\beta)$} \label{Sabd}

In his Comment \cite {Burde}, 
the author points out that the consistent second order perturbation approach can be achieved when all small parameters are related to only one, assuming for instance 
\begin{equation} \label{small}
\alpha= A\beta, \quad \delta = q\beta,
\end{equation}
where the constants $A,q$ are of the order of 1. 
The presence of the factors $A$ and $q$ in the following steps eases to recognize the origin of particular terms.

In standard approach the velocity potential is assumed in the form of the series $\phi(x,z,t)=\sum_{m=0}^{\infty} z^m \phi^{(m)}(x,t)$. For flat bottom case ($\delta=q=0$) equations (\ref{2BS}) and (\ref{6BS}) allow us to express all $\phi^{(m)}(x,t)$ with even $m$ only, by $f(x,t):=\phi^{(0)}(x,t)$ and its even $x$-derivatives. For the uneven bottom case, to satisfy the equation (\ref{6BS}), the velocity potential has to contain also odd $m$ terms. In general the velocity potential fulfiling Laplace equation can be expressed in the following form
\begin{align} \label{potG}
\phi(x,z,t) & =\sum_{m=0}^{\infty} \frac{(-1)^m \beta^m}{(2m)!}\frac{\partial^{2m} f}{\partial x^{2m}} z^{2m} \\ & + \sum_{m=0}^{\infty} \frac{(-1)^m \beta^{m+1}}{(2m+1)!}\frac{\partial^{2m+1} F}{\partial x^{2m+1}} z^{2m+1}, \nonumber \end{align}
where $F=F(x,t)$. Explicit form of this velocity potential is 
\begin{align} \label{potEx}
\phi & = f-\frac{1}{2}\beta z^2 f_{2x} + \frac{1}{24}\beta^2 z^4 f_{4x} - \frac{1}{720}\beta^3 z^6 f_{6x} + \cdots  \nonumber \\ & 
+ \beta z G -\frac{1}{6}\beta^2 z^3 G_{2x} + \frac{1}{120}\beta^3 z^5 G_{4x}+ \cdots ,
\end{align}
where $G=F_x$. 
Substituting (\ref{potEx}) into (\ref{6BS}) gives (with $z=q\beta h$) nontrivial relation between the functions $G$ and $f$ 
\begin{align} \label{GG}
G & -q\beta (h f_x)_x - q^2\beta^3 (h^2 G_x)_x +\frac{1}{6}q^3\beta^4 (h^3 f_{3x})_x \\ &
+\frac{1}{24}q^4 \beta^6 (h^4 G_{3x})_x-\frac{1}{120}q^5 \beta^7 (h^5 f_{5x})_x+ \cdots=0. \nonumber
\end{align}
To specify this relation let us express $G$ as a series
\begin{equation} \label{GGb}
G=G0 + \beta G1+ \beta^2 G2+ \beta^3 G3+ \beta^4 G4 +\cdots
\end{equation} 
Substituting (\ref{GGb}) into (\ref{GG}) and collecting powers of $\beta$ we get $G0=0,~ G2=0$ and
\begin{equation} \label{GGb1}
G1=q (h f_x)_x,~   G3=q^2 (h^2 G_x)_x ,~  G4=\frac{1}{6}q^3 (h^3 f_{3x})_x \hspace{1ex}\ldots
\end{equation} 
So, the function $G$ is given by
\begin{equation} \label{GGc}
G =\beta q (h f_x)_x + \beta^3  q^2 (h^2 G_x)_x + \ldots
\end{equation} 
Since we are interested in second order equations, we can safely reject all terms except the first one in (\ref{GGc}) since after substitution of (\ref{GGc}) to the velocity potential (\ref{potEx}) they contribute in at least the fourth order in $\beta$. This approximation allows us to express the $x$-dependence of the velocity potential through $f,h$ and their $x$-derivatives. 

\emph{Remark: The form of (\ref{GGc}) indicates that attempts to derive a wave equation of the order higher than second are practically unfeasible.}

Then we obtain velocity potential in the following form
\begin{align} \label{pot16}
\phi & = f-\frac{1}{2}\beta z^2 f_{2x} + \frac{1}{24}\beta^2 z^4 f_{4x} - \frac{1}{720}\beta^3 z^6 f_{6x} + \cdots  \nonumber \\ & 
+ \beta^2 z q (h f_x)_x -\frac{1}{6}\beta^3 z^3 q (h f_x)_{3x} + \frac{1}{120}\beta^4 z^5 q (h f_x)_{5x}  \nonumber\\ & +\cdots  
\end{align}
Inserting (\ref{pot16}) into (\ref{4BS}) and (\ref{5BS}) and retaining terms up to second order yields the set of the Boussinesq equations in the following form (as usual $w=f_x$) 
\begin{align} \label{Bus2}
 \eta_t + w_x & + \beta\left(A(\eta w)_x-\frac{1}{6} w_{3x} -q(hw)_x\right) \\
   &  +\beta^2\left(-A\frac{1}{2} (\eta w_{2x})_ x +\frac{1}{120}w_{5x} - q (hw)_{3x} \right) =0 \nonumber  \\ \label{Bus3} 
w_t + \eta_x & +\beta\left(w w_x -\frac{1}{2} w_{2xt}\right)  \\  & + 
\beta^2\left[A\left(-(\eta w_{xt})_x +\frac{1}{2} w_x w_{2x} - \frac{1}{2} w w_{3x}\right) \right. \nonumber \\ & \hspace{6ex} \left. + \frac{1}{24} w_{4xt}+q(hw_t)_{2x}\right] =0 .\nonumber
\end{align} 
Inserting $A\beta=\alpha$ and $q\beta=\delta$ into (\ref{Bus2})-(\ref{Bus3}) we regain Eqs.~(8)-(9) from \cite{KRI}, as well as  Eqs.~(8)-(9) in Section~\ref{intro}.

Below we prove that the Boussinesq set (\ref{Bus2})-(\ref{Bus3}) cannot be reduced to the KdV - type wave equation even in the first order. 
It is well known that in the lowest (zero) order the Boussinesq set reduces to 
\begin{align} \label{0ord}
\eta_t + w_x &=0, \quad   w_t + \eta_x =0.\\
 \Longrightarrow \hspace{8ex}  w &=\eta, \quad \eta_t + \eta_x =0. \nonumber
\end{align}
In the first order the Boussinesq set reduces to 
\begin{align} \label{4hx1}
  \eta_t + w_x  + \alpha(\eta w)_x-\frac{1}{6}\beta w_{3x}  -
 \delta (hw)_x & =0,  \\ \label{5hx1}
w_t + \eta_x + \alpha w w_x -\frac{1}{2}\beta\, w_{2xt} &  =0 . 
\end{align} 
Assume that in the first order 
\begin{equation} \label{w1o}
w= \eta + \alpha \left(-\frac{1}{4}\eta^2\right) + \beta \left(\frac{1}{3}\eta_{2x}\right) + \delta Q,
\end{equation}
since it is well known that for $\delta=0$, that is, for the flat bottom case, the form of two first corrections implies KdV equation.
Then we substitute (\ref{w1o}) into equations (\ref{4hx1})-(\ref{5hx1}), express time derivatives in terms of $x$-derivatives from zeroth order relations (\ref{0ord}) and retain term only to the first order. This yields
\begin{equation} \label{4hx2}
\eta_t+\eta_x+\alpha\frac{3}{2}\eta\eta_x+\beta\frac{1}{6}\eta_{3x}
 + \delta (Q_x- (h\eta)_x) =0
\end{equation} 
and
\begin{equation} \label{5hx2}
\eta_t+\eta_x+\alpha\frac{3}{2}\eta\eta_x+\beta\frac{1}{6}\eta_{3x}
 + \delta Q_t  =0 .
\end{equation}
Subtracting (\ref{5hx2}) from (\ref{4hx2}) one obtains the condition
\begin{equation} \label{Qc}
Q_x-Q_t =  (h\eta)_x.
\end{equation}
Since $Q$ has to be expressed by $h,\eta$ and possibly their derivatives, it is easy to see that $Q_t=-Q_x$ cannot be true for arbitrary bottom function $h(x)$. Therefore, for arbitrary bottom profile the Boussinesq set (\ref{4hx1})- (\ref{5hx1}) cannot be made compatible and the unidirectional wave equation of KdV-type cannot be derived. Then higher order wave equations cannot be derived, as well.

In \cite{Burde}, 
the author claimed that for the first order Boussinesq's equations 
(\ref{Bus2})-(\ref{Bus3}) the appropriate correction $Q$ can be found for the specific case $h(x) = kx$. He proposed 
\begin{equation} \label{wabd}
w = \eta-\alpha\frac{1}{4}\eta^2 +\beta \left(\frac{1}{3}\eta_{2x}+ q\frac{1}{4}(2kx\eta+k\!\! \int\!\! \eta dx)\! \right).
\end{equation} 
Insertion (\ref{wabd}) into (\ref{Bus2}) yields 
\begin{align} \label{Bus2kx}
\eta_t+\eta_x & +\alpha (2\eta\eta_x) \\ & +\beta\!\left(\!-\frac{1}{2}\eta\eta_x \!+\frac{1}{6}\eta_{3x}\! -\frac{1}{4} q k(\eta+2x\eta_x)\!\right)=0, \nonumber
\end{align}
whereas insertion (\ref{wabd}) into (\ref{Bus2}) gives (after replacing $t$-derivatives by $x$-derivatives)
\begin{align} \label{Bus3kx}
\eta_t+\eta_x +\beta\!\left(\!\frac{3}{2}\eta\eta_x \!+\frac{1}{6}\eta_{3x}\! -\frac{1}{4} q k(\eta+2x\eta_x)\!\right)=0. 
\end{align}
The equations (\ref{Bus2kx}) and (\ref{Bus3kx}) are compatible only for $\alpha=\beta$ (or equivalently for $A=1$ in notation used in \cite{Burde}).

The bottom function $h(x)=kx$ is unbound on $ x \in \mathbb {R} $ which contradicts the definition of the parameter $\delta = \frac{a_h}{h} $, where $ a_h $ is the amplitude of the bottom function. Also from a physics standpoint, the bottom function can not grow infinitely, because for some values of $ x $ the bottom would be above the water surface.

In \cite{KRnody} we showed that the class of bottom functions for which the Boussinesq equations (\ref{Bus2})-(\ref{Bus3}) can be made compatible is wider than the linear functiom $h=kx$. It is sufficient that $h_{2x}=0$, so $h(x)$ can be an arbitrary piecewise linear function. For such a function, the amplitude of bottom changes can be small everywhere. 
If the condition $h_{2x}=0$ is fulfiled, then the compatibility condition (\ref{Qc}) is satisfied by 
\begin{equation} \label{small2}
Q=\frac{1}{4}\left( h\eta+ h_x \! \int\! \eta \,dx\right).
\end{equation}
With this correction term the resulting wave equation, generalizing KdV equation for a piecewise linear bottom takes the following form
\begin{align} \label{kdvPL}
\eta_t+\eta_x +\frac{3}{2}\eta\eta_x +\frac{1}{6}\beta\eta_{3x} -\frac{1}{4}
\delta\left(2 h\eta_x + h_x\eta \right)=0. 
\end{align}
 In Section \ref{secN} we examine this case in numerical simulation.

\section{Derivation of the nonlinear wave equation for the case of  \boldmath$\alpha=O(\beta)$ and  \boldmath$\delta=O(\beta^2)$} \label{Sabd2}

In this case we set 
\begin{equation} \label{small2}
\alpha= A\beta, \quad \delta = q\beta^2.
\end{equation}
Now, we insert the general form of velocity potential (\ref{potEx}) into the bottom boundary condition which in this case is
\begin{equation} \label{bbc}
\phi_z-q\beta^3\left( h_x\,\phi_x\right) =  0,\quad\mbox{for}\quad z=q\beta^2 h(x) 
\end{equation}
obtaining relation similar to (\ref{GG})
\begin{align} \label{GG2}
G & -q\beta^2 (h f_x)_x -\frac{1}{2} q^2\beta^5 (h^2 G_x)_x +\frac{1}{6}q^3\beta^7 (h^3 f_{3x})_x \\ &
+\frac{1}{24}q^4 \beta^{10} (h^4 G_{3x})_x-\frac{1}{120}q^5 \beta^{12} (h^5 f_{5x})_x+ \cdots=0. \nonumber
\end{align}
Then, in the lowest order 
\begin{equation} \label{Gd2}
G = q\beta^2 (h f_x)_x
\end{equation}
which inserted into (\ref{potEx}) gives the velocity potential as
\begin{align} \label{pot16a}
& \phi = f-\frac{1}{2}\beta z^2 f_{2x} + \frac{1}{24}\beta^2 z^4 f_{4x} - \frac{1}{720}\beta^3 z^6 f_{6x} + \cdots   \\ & 
+\! q \beta^3 z (h f_x)_x \!-\!\frac{1}{6}q \beta^4 z^3 (h f_x)_{3x}\! +\! \frac{1}{120} q\beta^5 z^5 (h f_x)_{5x}\!+\! \cdots ,\nonumber
\end{align}

In this case the Boussinesq system has the form
\begin{align} \label{4HX}
  \eta_t + w_x &+   \beta\left(A(\eta w)_x-\frac{1}{6} w_{3x}\right)  \\
  & +\beta^2\left(-A\frac{1}{2} (\eta w_{2x})_ x +\frac{1}{120}w_{5x} - q\,(hw)_x\right)  =0,\nonumber  \\ \label{5HX}
 w_t + \eta_x & +\beta\left(A w w_x -\frac{1}{2} w_{2xt}\right) \\ & +  \beta^2\left(-A(\eta w_{xt})_x +A\frac{1}{2} w_x w_{2x} - A\frac{1}{2} w w_{3x}\right.\nonumber\\
& \hspace{7ex} \left.
+ \frac{1}{24} w_{4xt} \right) =0.\nonumber
\end{align} 

In the first order this system reduces to the common KdV system, with 
\begin{equation} \label{h0}
w =  \eta +   \beta\left(-A\frac{1}{4}\eta^2 +\frac{1}{3}\eta_{2x} \right) 
\end{equation}
which ensures the KdV equation
\begin{equation} \label{kdv1}
\eta_t + \eta_x +\beta\left(A\frac{3}{2}\eta\eta_x+ \frac{1}{6}\eta_{3x} \right)=0 .\end{equation}

Now, we aim to satisfy the Boussinesq system (\ref{4HX})-(\ref{5HX}) with the terms of the second order included. 
 Then, we set (the first term with $\beta^2$ makes the set (\ref{4HX})-(\ref{5HX}) compatible for the flat bottom case)
\begin{align} \label{wab2}
w = \eta & +\beta\left(-A\frac{1}{4}\eta^2 +\frac{1}{3}\eta_{2x} \right) \\ &
 + \beta^2 \left(A^2\frac{1}{8}\eta^3 +A\frac{3}{16}\eta_x^2 +A\frac{1}{2}\eta \eta_{2x}+\frac{1}{12}\eta_{4x} \right)\nonumber \\ & 
+ \beta^2 q\, Q.\nonumber
\end{align}
Next, we insert the trial function (\ref{wab2}) into (\ref{4HX}) and (\ref{5HX}) and retain terms up to second order in $\beta$. Proceeding analogously as in the case of first order we find that compatibility of the Boussinesq equations (\ref{4HX})-(\ref{5HX})  requires the same condition (\ref{Qc}) for the correction function $Q$
$$ Q_x- Q_t =  (h \eta)_x.$$
\indent
Note, that in order to replace $t$-derivatives by $x$-derivatives one has to use  the properties of the first order equation (\ref{kdv1}), that is,  $ \eta_t= -\eta_x-\beta\left( A\frac{3}{2}\eta\eta_x +\frac{1}{6}\eta_{3x}\right)$ and its derivatives. \\
\indent
Using the formula for the correction functions (\ref{Qc}),  for a piecewise linear bottom,  
we obtained in this case, $\alpha=O(\beta), \delta=O(\beta^2)$, the equation 
\begin{align}\label{2abQ}
\eta_t+\eta_x &+\frac{3}{2}\alpha\eta\eta_x  +\frac{1}{6}\beta\eta_{3x} 
-\frac{3}{8}\alpha^2\eta^2\eta_x \\ & 
+\alpha\beta\left(\frac{23}{24}\eta_x^2  + \frac{5}{12}\eta\eta_{2x} \right)\nonumber   \\ & + \beta^2 \left(\frac{19}{360}\eta_{5x} \right) -\frac{1}{4}\delta (2 h \eta_x+h_x \eta) =0
\nonumber  
\end{align}
which generalizes the extended KdV (KdV2) equation (\ref{etaabd}) for piecewise linear bottom profiles.

These forms of equation (\ref{2abQ}) may be misleading, since the terms with $\delta$, looking as first order ones, are, in fact, of second order.

The equation (\ref{2abQ}), limited to the case $\delta=q=0$,  is the \emph{extended KdV equation or KdV2}  \cite{MS90}. 
This equation is nonintegrable. Despite this fact, we found several forms of analytic solutions to KdV2: soliton solutions in \cite{KRI}, cnoidal solutions ($\sim \text{cn}^2$) in \cite{IKRR} and superposition cnoidal solutions ($\sim \text{dn}^2\pm \sqrt{m}\,\text{cn\,dn}$) in \cite{RKI,RK}. 

The wave equation (\ref{2abQ}) is very similar to the erroneous \cite[Eq.~(18)]{KRI}. The latter contains, apart from the leading term from the bottom $-\frac{1}{4}\delta(2 h \eta_x+h_x \eta) $, two other terms which resulted from not fully consistent derivation.

\section{Numerical tests} \label{secN}

In this section, we tentatively examine the motion of appropriate solitons entering the region where the bottom is no longer even. In these tests, we use our numerical code based on the finite difference method. The code was described in detail in \cite{KRI}.

\subsection{The case of  \boldmath$\alpha=O(\beta)$ and  \boldmath$\delta=O(\beta)$}

In this part we present evolution of the KdV solitons obtained with numerical solution of the equation (\ref{Bus3kx}). Since this equation is valid only for $\alpha=\beta$ and all three parameters should be of the same order we set 
in these test $\alpha=\beta=\delta=0.25$. As a bottom function $h(x$) we chose a piecewise function of trapezoid shape located at $x_1=5, x_2=10, x_3=20, x_4=25$. Since the equation (\ref{kdvPL}) is valid only for the piecewise  linear bottom function $h(x)=kx$ the trapezoidal bottom is allowed. The size and location of the trapezoid allows us also to compare the results with those presented in \cite{RRIK}. Note that the bottom function is drawn not in scale. The initial condition is the KdV soliton with the amplitude equal to 1, that is, $\eta(x,t=0)=\text{sec}^2\left(\sqrt{\frac{3\alpha}{4\beta}}\,x\right)=\text{sec}^2\left(\sqrt{0.75}\,x\right)$.

\begin{figure}[htb]
\begin{center}
 \resizebox{1.01\columnwidth}{!}{\includegraphics[angle=270]{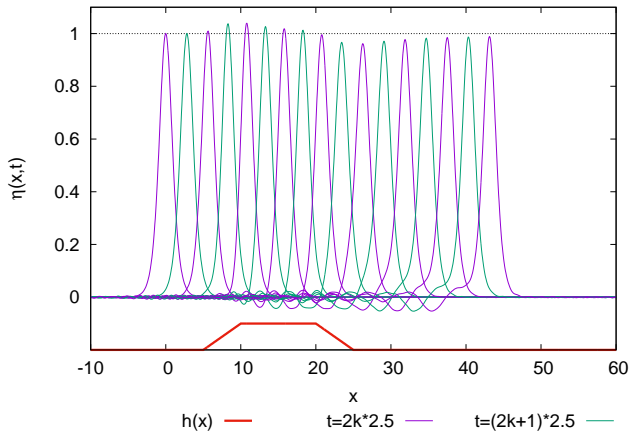}}
 \caption{Time evolution of the KdV soliton entering the trapezoidal bump, $t\in [0,30]$. } \label{abd025U}
\end{center} 
 \end{figure}
\begin{figure}[htb]
\begin{center}
 \resizebox{1.01\columnwidth}{!}{\includegraphics[angle=270]{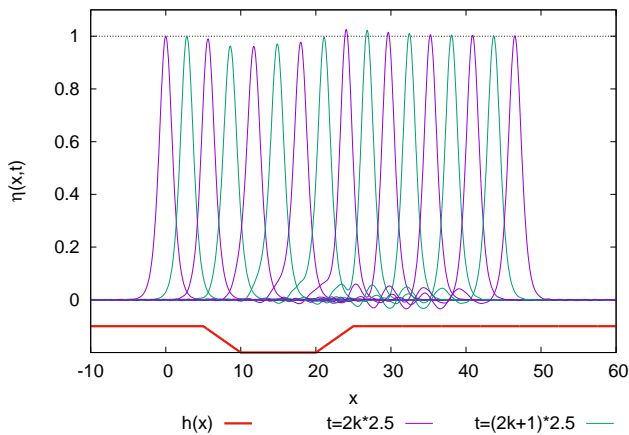}}
 \caption{Time evolution of the KdV soliton entering the trapezoidal well, $t\in [0,30]$.} \label{abd025D}
\end{center} 
 \end{figure}

In the case presented in Fig.~\ref{abd025U} the soliton first slows down and then accelerates with the amplitude increase and decrease, respectively.
In the case presented in Fig.~\ref{abd025D} the soliton first accelerates and then slows down with the amplitude decrease and increase, respectively.
Therefore in the latter case, the distance covered by the soliton at $t=40$ is larger than in the former case.

In both cases, the interaction with the uneven bottom produces additional wave trains with small amplitudes behind the main wave.

\subsection{The case of \boldmath$\alpha=O(\beta)$ and \boldmath$\delta=O(\beta^2)$}

In this case there exists KdV2 solitons, that is solitons of the equation (\ref{etaabd}), that is, for the flat bottom (see, Sect.~V in \cite{KRI}). The amplitude of such solitons is one for $\alpha\approx 0.2424$. Since we compare motion of solitons with the same amplitude (equal to 1) we present below numerical solutions when $\alpha=\beta=0.2424$ and $\delta=2\beta^2\approx 0.1175$. Then the initial condition is  $\eta(x,t=0)=\text{sec}^2\left(\sqrt{0.599}\,x\right)$.

\begin{figure}[htb]
\begin{center}
 \resizebox{1.01\columnwidth}{!}{\includegraphics[angle=270]{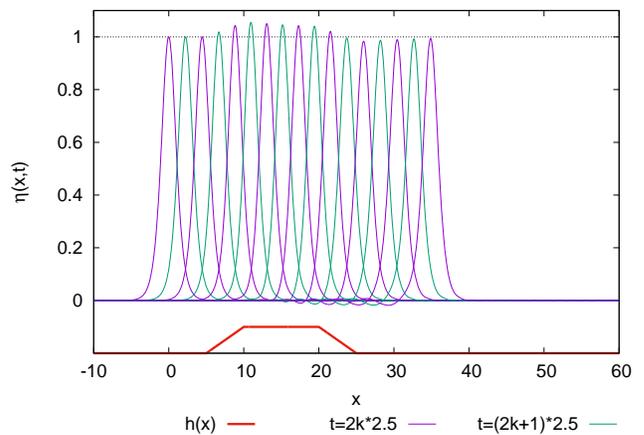}}
 \caption{Time evolution of the KdV2 soliton entering the trapezoidal bump, $t\in [0,30]$. } \label{ab024dU}
\end{center} 
 \end{figure}

\begin{figure}[htb]
\begin{center}
 \resizebox{1.01\columnwidth}{!}{\includegraphics[angle=270]{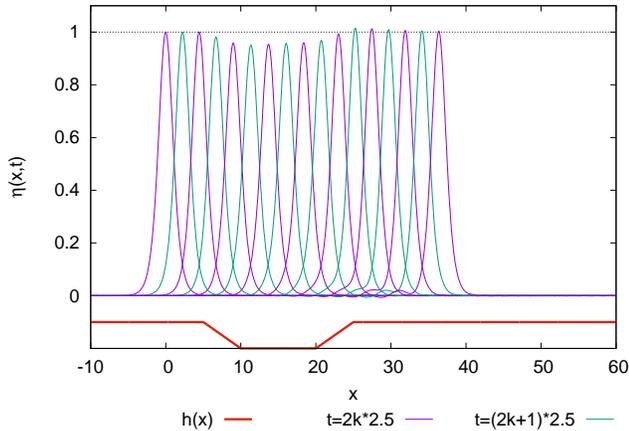}}
 \caption{Time evolution of the KdV2 soliton entering the trapezoidal well, $t\in [0,30]$.} \label{ab024dD}
\end{center} 
 \end{figure}

Distortions of the soliton shape caused by interaction with uneven bottom observed in Figs.~\ref{ab024dU} and \ref{ab024dD} are much smaller than those in Figs.~\ref{abd025U} and \ref{abd025D}.

Comparison of the numerical evolution of KdV2 solitons obtained with the equation (\ref{2abQ}) with that resulted from the erroneous equation \cite[Eq.~(18)]{KRI} shows an important difference. The radiation of small amplitude wavetrain in front of the main wave, present in evolution according to \cite[Eq.~(18)]{KRI} seems to dissappear in evolution according to the equation (\ref{2abQ}) displayed in Figs.~\ref{ab024dU} and~\ref{ab024dD}. 

\begin{figure}[htb]
\begin{center}
 \resizebox{1.01\columnwidth}{!}{\includegraphics[angle=270]{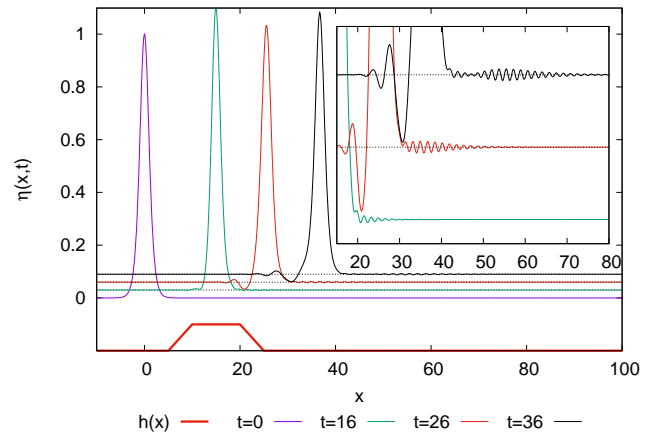}}
 \caption{
Several profiles of the KdV2 soliton moving over the trapezoidal well. In the insert the radiation of the faster wave packets of small amlitude is clearly seen. } \label{ab024d0176U}
\end{center} 
\end{figure}

The thorough inspection of the calculated data reveals that this radiation still exists, but with much smaller amplitude (in Figs.~\ref{ab024dU} and \ref{ab024dD} this amplitude is comparable to the linewidth). In order to enhance this effects we performed additional calculations in which we set $\alpha=\beta=0.2424$ and $\delta=3\beta^2\approx 0.176$. Several profiles of the wave obtained in the numerical evolution of KdV2 soliton according to the equation (\ref{2abQ}) are displayed in Fig.~\ref{ab024d0176U}.

 The creation and then detachment of the small amplitude wave packet in front of the main wave is clearly exposed in the insert. This is qualitatively the same feature as observed in our previous papers \cite{KRI,KRR,RRIK} for wave motion according to the erroneous equation \cite[Eq.~(18)]{KRI}. Quantitatively the effect has much smaller amplitude, for realistic values of parameters $\alpha,\beta,\delta$ it is smaller than 1\% of the solitons amplitude.
On the other hand, even such small effect suggests the origin of the very tiny wrinkles observed always on the water surface at the seashore.

We are sure that this is the real effect, not an artifact of numerical simulation. Since our code utilizes periodic boundary conditions we performed calculations on much wider $x$-interval than displayed in figures above. In such cases, when the soliton moves far from the end of the $x$-interval, the boundary conditions do not influence the shape of the localized wave.

\begin{acknowledgments}
We thank the author of \cite{Burde} for his detailed explanation of consistent perturbation approach in the case of several small parameters. His remarks allowed us to derive the correct second order wave equation explicitly containing terms from the uneven bottom, for arbitrary, bounded bottom function. 
\end{acknowledgments}



\begin{thebibliography}{99}

\bibitem{KRR} Karczewska, A., Rozmej, P. and Rutkowski, \L{}.:
\emph{A new nonlinear equation in the shallow water wave problem},
Physica Scripta, \textbf{89}, 054026, (2014).  
DOI: 10.1088/0031-8949/89/5/054026

\bibitem{KRI} Karczewska, A., Rozmej, P. and Infeld, E.:
\emph{Shallow water soliton dynamics beyond KdV},
Physical Review E, \textbf{90}, 012907, (2014). 
DOI: 10.1103/PhysRevE.90.012907

\bibitem{kdv} Korteweg, D.J. and de Vries, G.: \emph{On the change of form of the long waves advancing in a rectangular canal, and on a new type of stationary waves}, 
Phil.\ Mag.\ (5), {\bf 39}, 422 (1895). 

\bibitem{MS90} Marchant, T.R. and Smyth, N.F.:
\emph{The extended Korteweg-de Vries equation and the resonant flow
of a fluid over topography}, Journal of Fluid Mechanics, \textbf{221}, 263-288, (1990).

\bibitem{Burde} \emph{Comment on “Shallow-water soliton dynamics beyond the Korteweg–de Vries equation” by A.\ Karczewska, P.\ Rozmej and E.\ Infeld, Phys.\ Rev.\ E 90, 012907 (2014)} manuscript written by the anonymous referee. This manuscript was used as justification of rejection of our paper, P.~Rozmej, A.~Karczewska, \emph{Comment on the paper "The third-order perturbed Korteweg-de Vries equation for shallow water waves with a non-flat bottom" by M. Fokou, T.C. Kofané, A. Mohamadou and E. Yomba, Eur. Phys. J. Plus, 132, 410 (2017)}, arXiv:1804.01940, submitted by us to The European Physical Journal Plus.


\bibitem{KRnody} Karczewska, A. and Rozmej,~P.:
\emph{What kinds of KdV-type equations are allowed by an uneven bottom}.
arXiv:1903.04890.

\bibitem{IKRR}  Infeld, E., Karczewska, A., Rowlands, G. and Rozmej,~P.:  \emph{Exact cnoidal solutions of the extended KdV equation}, Acta Phys. Pol. A, \textbf{133}, 1191-1199, (2018). 
DOI: 10.12693/APhysPolA.133.1191

\bibitem{RKI} Rozmej, P., Karczewska, A. and Infeld, E.: \emph{Superposition solutions to the extended KdV equation for water surface waves}, Nonlinear Dynamics  \textbf{91}, 1085-1093, (2018). 
DOI: 10.1007/s11071-017-3931-1

\bibitem{RK} Rozmej, P. and Karczewska, A.:
\emph{New Exact Superposition Solutions to KdV2 Equation},  Advances in Mathematical Physics. \textbf{2018}, Article ID 5095482, 1-9, (2018).  
DOI: 10.1155/2018/5095482

\bibitem{RRIK}  Rowlands, G., Rozmej, P., Infeld, E. and Karczewska, A.:
\emph{Single soliton solution to the extended KdV equation over uneven depth}, Eur. Phys. J. E, \textbf{40},  100, (2017). 
DOI: 10.1140/epje/i2017-11591-7

\end{thebibliography}
\end{document}